\documentclass[twocolumn,preprintnumbers,superscriptaddress,nofootinbib,aps,prd,floatfix]{revtex4}

\usepackage{enumerate}
\usepackage{amsmath,amssymb}
\usepackage{graphicx}
\usepackage{bbm,slashed}
\usepackage{bbold}
\usepackage{xspace,slashed}
\usepackage{hyperref}
\hypersetup{colorlinks=true, citecolor=blue, urlcolor=blue, linkcolor=blue}
\usepackage[normalem]{ulem}
\usepackage{subfigure}
\usepackage{multirow,array}
\usepackage{comment}
\usepackage{float}

\usepackage{tabularx,booktabs} 
\newcolumntype{C}{>{$}c<{$}}
\AtBeginDocument{%
  \heavyrulewidth=.08em
  \lightrulewidth=.05em
  \cmidrulewidth=.03em
  \belowrulesep=.65ex
  \belowbottomsep=0pt
  \aboverulesep=.4ex
  \abovetopsep=0pt
  \cmidrulesep=\doublerulesep
  \cmidrulekern=.5em
  \defaultaddspace=.5em
}

\begin{document}

\providecommand{\abs}[1]{\lvert#1\rvert}

\newcommand{\Znunujets}{(Z\to{\nu\bar{\nu}})+\text{jets}}
\newcommand{\Welnujets}{(W\to{\ell\nu})+\text{jets}}
\newcommand{\Znunujet}{(Z\to{\nu\bar{\nu}})+\text{jet}}
\newcommand{\Welnujet}{(W\to{\ell\nu})+\text{jet}}

\title{Non-linear gauge-Higgs CP violation}
\begin{abstract}
\noindent A critical element of the LHC physics program is the search for an additional source of CP violation. This is largely unexplored in the context of non-linear Higgs physics, which is naturally described in Higgs Effective Field Theory (HEFT). Relevant new higher-dimensional operators modify the production rate and branching ratios of the Higgs boson, de-correlating different Higgs multiplicities. In this work, we consider single Higgs and Higgs pair production via weak boson fusion from the perspective of gauge-Higgs CP violation through the lens of Higgs non-linearity. This generalizes existing rate-based searches and analyses by the ATLAS and CMS experiments. Particular focus is given to the phenomenological differences in the expected BSM sensitivity pattern when comparing HEFT constraints with Standard Model Effective Field Theory (SMEFT) limits. 
\end{abstract}

\author{Akanksha Bhardwaj}\email{akanksha.bhardwaj@okstate.edu}
\affiliation{Department of Physics, Oklahoma State University, Stillwater, OK, 74078, USA\\[0.1cm]}
\author{Christoph Englert} \email{christoph.englert@glasgow.ac.uk}
\affiliation{School of Physics \& Astronomy, University of Glasgow, Glasgow G12 8QQ, UK\\[0.1cm]}
\author{Dorival Gon\c{c}alves}\email{dorival@okstate.edu} 
\affiliation{Department of Physics, Oklahoma State University, Stillwater, OK, 74078, USA\\[0.1cm]}
\author{Alberto Navarro}\email{alberto.navarro\_serratos@okstate.edu}
\affiliation{Department of Physics, Oklahoma State University, Stillwater, OK, 74078, USA\\[0.1cm]}

\maketitle

\allowdisplaybreaks

\section{Introduction}
\label{sec:intro}
\noindent The search for new physics beyond the Standard Model (BSM) has so far been unsuccessful. The lack of concrete (\emph{i.e.} resonant) evidence therefore typically serves as a motivation to consider a large mass gap between the spectrum of the Standard Model (SM) and its ultraviolet (UV) completion~\cite{PhysRevD.11.2856}. Under these assumptions, employing Effective Field Theory (EFT) methods is well-motivated. This methodology has a long-standing tradition in physics and has seen rapid progress in its application to the experimental program at the Large Hadron Collider (LHC). EFT, when derived from the symmetry pattern of the SM (dubbed SMEFT)~\cite{Grzadkowski:2010es} and its relation to more general parameterisations highlighting the custodial isosinglet nature of the Higgs boson (referred to as Higgs Effective Field Theory, or HEFT) are topics of recent theoretical and phenomenological interest~\cite{Brivio:2013pma,Brivio:2016fzo}. 

On the one hand, a more general parametrization than SMEFT of Higgs interactions covers a wider class of models for UV matching~\cite{Alonso:2016oah,Brivio:2017vri,Gomez-Ambrosio:2022qsi,Gomez-Ambrosio:2022why,Salas-Bernardez:2022hqv,Dawson:2023ebe,Dawson:2023oce,Delgado:2023ynh}, often with faster convergence in the EFT expansion~\cite{Helset:2020yio}. SMEFT correlations can be recovered from HEFT interactions through appropriate parameter choices and field redefinitions. On the other hand, SMEFT exclusion constraints are driven by the correlations imparted by assuming the Higgs as part of a weak doublet. Although the current Higgs measurements, so far, are largely compatible with a doublet-like character of electroweak symmetry breaking, this could downplay the new physics potential of rare final states that are becoming increasingly accessible towards the high-luminosity (HL) LHC phase. In this note, we take these observations as motivation to revisit CP violation in the gauge-Higgs sector from the perspective of Higgs non-linearity. For a corresponding study on CP violation in the fermion-Higgs sector, see Ref.~\cite{Bhardwaj:2023ufl}. 

BSM Higgs boson interactions can introduce additional sources of CP violation, which can address one of the Sakharov conditions~\cite{Sakharov} that the SM does not fulfill~\cite{Basler:2021kgq,Anisha:2022hgv,Goncalves:2023svb}. Current analyses at the LHC focus on single Higgs physics in the $ZZ$, $WW$ and $\gamma\gamma$ final states, but there is increasing progress in gaining sensitivity to rare processes such as weak boson fusion (WBF) Higgs pair production $pp\to hhjj$~\cite{ATLAS:2023qzf,CMS:2022hgz}. This latter process has interesting properties in the SM and beyond, as it directly probes geometric aspects of electroweak symmetry breaking~\cite{Alonso:2016oah,Alonso:2021rac,Englert:2023uug}. The phenomenology of this process is governed by SM unitarity identities, making it a formidable tool to discern the properties of electroweak symmetry breaking. As more data become available, WBF Higgs pair production will also be a natural playground to constrain Higgs boson non-linearity. The additional aspect of CP violation considered here extends the current searches performed by the experiments~\cite{ATLAS:2023qzf,CMS:2022hgz} in these channels. 

It is currently unclear whether angular (\emph{i.e.} dijet~\cite{Plehn:2001nj}) correlations can be isolated in WBF $hh$ production -- these serve as tools for comparing the correlated CP properties of the trilinear $\sim V^2h$ and quartic $\sim V^2h^2$ Higgs interactions ($V=W,Z,\gamma$). However, the rate constraints from Ref.~\cite{ATLAS:2023qzf,CMS:2022hgz} can be generalized to include aspects of CP violation, as we discuss in this work.

This paper is organized as follows: In Section~\ref{sec:corr}, we highlight the differences between HEFT and SMEFT expectations. We focus on SMEFT interactions in the gauge-Higgs sector that have direct counterparts in HEFT, except for Higgs multiplicity considerations. This enables us to systematically comment on the susceptibility of SMEFT constraints to widening correlations. In Sec.~\ref{sec:analysis}, we present the HL-LHC  sensitivity for HEFT and SMEFT, with a special focus on the phenomenology of the WBF $hh$ channel. We conclude in Sec.~\ref{sec:conclusion}.

\section{SMEFT and HEFT Correlations}
\label{sec:corr}
\noindent The SMEFT is constructed by assuming the Higgs field as an $SU(2)_L$ doublet~\cite{Buchmuller:1985jz,Hagiwara:1986vm,Grzadkowski:2010es}. The new interactions are parametrized through an expansion in higher dimensional operators, which are invariant under the Lorentz and SM gauge symmetries
\begin{equation}
\mathcal{L} = \mathcal{L}_{\text{SM}} + \sum_i \frac{c_i}{\Lambda^2} \mathcal{O}_i\,,
\end{equation}
 where  $\mathcal{L}_{\text{SM}}$ is the SM Lagrangian, ${c_i}$ are the Wilson coefficients, and $\Lambda$ is the scale of new physics.

The CP-even SMEFT operators relevant for the interactions between gauge bosons and the Higgs field are 
\begin{equation}
\label{eq:ops1}
\begin{split}
    {\cal{O}}_{\Phi B} &= {c_{\Phi B}\over \Lambda^2}  \Phi^\dagger \Phi  B^{\mu\nu}B_{\mu\nu}\,,\\
  {\cal{O}}_{\Phi W} &=  {c_{\Phi W}\over \Lambda^2}\Phi^\dagger \Phi  W^{i\,\mu\nu}W^{i}_{\mu\nu}\,, \\
  {\cal{O}}_{\Phi WB} &= {c_{\Phi WB}\over \Lambda^2} \Phi^\dagger \sigma^i W^{i\,\mu\nu}\Phi B_{\mu\nu}\,,
\end{split}
\end{equation}
where $\Phi$ represents the $SU(2)_L$ Higgs doublet, $W^\mu$ and $B^\mu$ denote the fields in the $SU(2)_L\times U(1)_Y$ gauge-field eigenbasis, and $\sigma^i$ are the Pauli matrices with $i=1,2,3$. The $S$ parameter severely constrains $c_{\Phi WB}$ and to discuss the (non-)linear sensitivity reach of WBF di-Higgs production, we set $c_{\Phi WB}=0$ in the following.\footnote{For a discussion on models that source this interaction pattern, see~\cite{Bakshi:2021ofj}. Note as the corresponding operator in HEFT is not correlated with the electroweak vacuum it is a priori unconstrained. For comparability, we will not consider it in the following.}
The CP-violating SMEFT operators affecting the gauge-Higgs boson interactions are given by 
\begin{equation}
\label{eq:ops2}
\begin{split}
    {\cal{O}}_{\Phi \widetilde{B}} &= {c_{\Phi\widetilde{B}}\over \Lambda^2}\Phi^\dagger \Phi  B^{\mu\nu}\widetilde{B}_{\mu\nu}\,,\\
  {\cal{O}}_{\Phi \widetilde{W}} &=  {c_{\Phi \widetilde{W}}\over \Lambda^2}\Phi^\dagger \Phi  W^{i\,\mu\nu}\widetilde{W}^{i}_{\mu\nu}\,, \\
  {\cal{O}}_{\Phi \widetilde{W}B} &= {c_{\Phi \widetilde{W}B}\over \Lambda^2} \Phi^\dagger \sigma^i  \Phi \widetilde{W}^{i\,\mu\nu}B_{\mu\nu}\,,
\end{split}
\end{equation}
where the dual field strength tensors are defined as $\widetilde X^{\mu\nu}=\epsilon^{\mu\nu\rho\delta}X_{\rho\delta}/2$. The operators of Eqs.~\eqref{eq:ops1} and~\eqref{eq:ops1} can form a closed set under the RGE flow~\cite{Jenkins:2013wua}, and studying them in isolation is justified. The phenomenological impact of these mixing effects is known to be small~\cite{Englert:2014cva}.

In the context of HEFT, the physical Higgs field $h$ and the three electroweak Goldstone bosons (GBs) $\pi^i$ are regarded as independent and not part of a $SU(2)$ doublet. In this scenario, the GBs are parametrized by a dimensionless unitary matrix $U(\pi)$~\cite{Coleman:1969sm,Callan:1969sn}
\begin{equation}
U(\pi) =\exp\left({i\over 2v}  \sigma^i\pi^i  \right)\,,
\end{equation}
where $v=246$~GeV is fixed by, {\it e.g.}, the $W$ mass. The matrix $U(\pi^i)$ transforms as a bi-doublet of the global symmetry $SU(2)_L \times SU(2)_R$, and its covariant derivative is given by
\begin{equation}
D_\mu U = \partial_\mu U + ig 
W^i_\mu \frac{\sigma^i}{2} 
U  + ig'B_\mu U {\sigma^3\over 2} .
\end{equation}
The leading order HEFT Lagrangian leads to $\kappa$-framework~\cite{LHCHiggsCrossSectionWorkingGroup:2011wcg} SM gauge-Higgs interactions via
\begin{subequations}
	\label{eq:heftlo}
	\begin{equation}
		\mathcal{L}^{\text{HEFT}}_{\text{LO}} = 
		 \frac{ v^2}{4} \mathcal{F}_{h} \,\text{Tr}[D_{\mu} U^{\dagger} D^{\mu} U]  \,.
 \end{equation}
The interactions of the (iso-)singlet Higgs field with gauge and Goldstone bosons  are parametrized by a polynomial function $\mathcal{F}_{h}$, written as
\begin{equation}
	 \label{eq:flare}
		\mathcal{F}_{h} = 1+ 2(1+\zeta_{1})\frac{h}{v} + 
		 (1+\zeta_{2}) \left(\frac{h}{v}\right)^2  + \dots\,.
\end{equation}
\end{subequations}
The choice of $\zeta_{1,2} = 0$ corresponds to the SM gauge-Higgs interactions. For phenomenological analyses, it is therefore convenient to isolate the new physics components $\zeta_1$ and $\zeta_2$. The Higgs boson couplings in this framework become uncorrelated free parameters. 

\begin{figure*}[t!]
   \centering
   \subfigure[]{\includegraphics[width=0.45\textwidth]{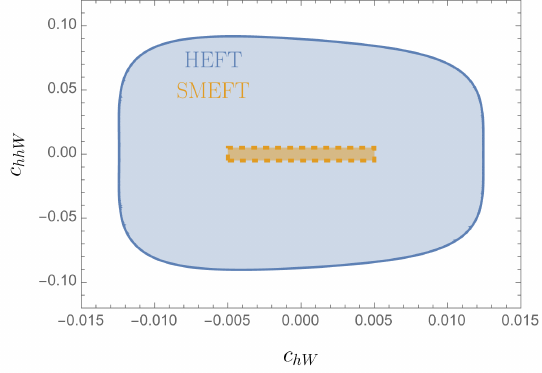}}
   \subfigure[]{\includegraphics[width=0.45\textwidth]{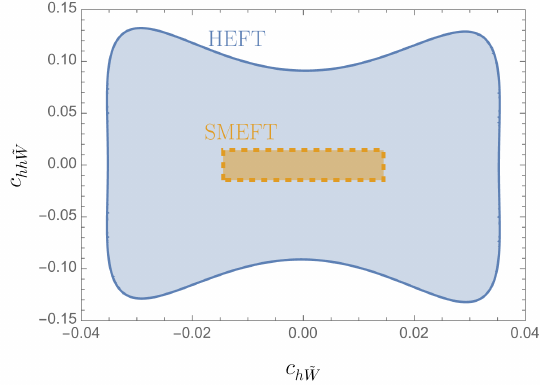}}
   \subfigure[]{\includegraphics[width=0.45\textwidth]{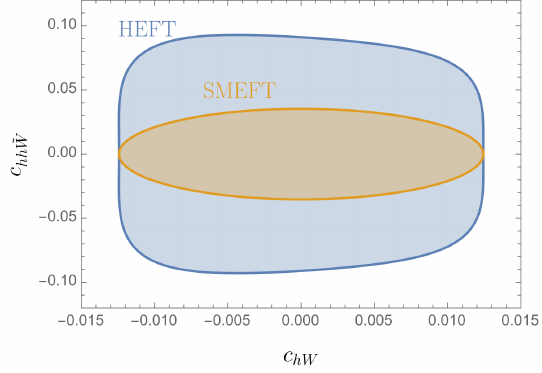}}
   \subfigure[]{ \includegraphics[width=0.45\textwidth]{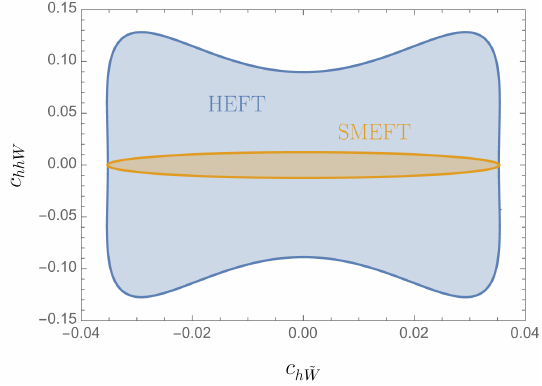}}
    \caption{$95\%$ confidence level regions for the Wilson coefficients: (a)  $(c_{hW},c_{hhW})$, (b) $(c_{h\tilde{W}},c_{hh\tilde{W}})$, (c)  $(c_{hW},c_{hh\tilde{W}})$, and (d) $(c_{h\tilde{W}},c_{hhW})$ in the HEFT (blue) and SMEFT (orange). These bounds are derived from single Higgs signal strength measurements from the CMS analysis~\cite{CMS:2021nnc}, which provide the leading sensitivities for the SMEFT framework. Additionally, the analysis incorporates the WBF di-Higgs channel to account for possible non-linearities. To simplify the comparison between HEFT and SMEFT, we set $\Lambda=v$. We consider the HL-LHC with $3~\text{ab}^{-1}$ of data. As the di-Higgs interactions are constrained by model-assumptions we present them as boxed contours (with dashed outline contours) to highlight this systematic difference in comparison with HEFT.}
       \label{fig:HEFTcorW}
\end{figure*}

The terms in Eq.~\eqref{eq:heftlo} do not lead to gauge-Higgs CP-violation and relevant interactions in HEFT arise at ${\cal{O}}(p^4)$ in the momentum expansion. Furthermore, the different Lorentz structures of $\zeta_1$ vs.~dimension six SMEFT operators can, in principle, be established from $gg\to h \to ZZ^\ast$ measurements, which will further inform \hbox{(multi-)}Higgs measurements in the WBF channel~\cite{Anisha:2024xxc}. To gauge the extent to which WBF can then probe non-linear deformations of the SM, we will set $\zeta_1=\zeta_2=0$ and consider the HEFT generalisation of the Lorentz structures related to the SMEFT operators of Eqs.~\eqref{eq:ops1} and~\eqref{eq:ops2} in the following. 

For the sake of defining our notation for HEFT and SMEFT couplings, let us consider the $B^{\mu\nu}B_{\mu \nu}$ contributions. In the language of HEFT, a series of interactions emerges due to the singlet nature of the Higgs field
\begin{multline}
{\mathcal{F}}_{h,B} {\text{Tr}}\left[ B_{\mu \nu} \sigma^3 B^{\mu \nu} \sigma^3 \right] \\=
\left( c_{hB} {h\over v} + c_{hhB} {h^2 \over 2 v^2} + ...\right)  B^{\mu\nu}B_{\mu\nu} \,.
\end{multline}
In HEFT, the  $VVh$ and $VVhh$  gauge-Higgs vertices originate from independent coefficients $c_{hB}$ and $c_{hhB}$, respectively. In contrast, in SMEFT these two interactions are governed by the same Wilson coefficient $c_{\Phi B}$, as they both correspond to the same SMEFT operator. Hence, the correspondence between HEFT and SMEFT is given by
\begin{align}
\label{eq:smeftheft}
c_{hB}\over v  &= c_{\Phi B} {v\over \Lambda^2}\,,
&c_{hhB} &= c_{\Phi B} {v^2\over \Lambda^2}\,,\\
c_{h\widetilde B}\over v &= c_{\Phi \widetilde B} {v \over \Lambda^2}\,, 
&c_{hh\widetilde B} &= c_{\Phi \widetilde B} {v^2\over \Lambda^2}\,.
\label{eq:smeftheft2}
\end{align}
Below in Sec.~\ref{sec:analysis}, we present the results of our analysis for both HEFT and SMEFT, using $\text{HEFT}\to \text{SMEFT}$ identifications. This will enable us to clarify to what extent  $\text{SMEFT}\subset\text{HEFT}$ is probable at the HL-LHC.

To formulate constraints on the BSM parameters, we can devise observables from the BSM coupling-expanded matrix elements. The scattering amplitude in the presence of higher-order terms can be written as the sum of SM ($\cal{M}_{\text{SM}}$) and BSM (${\cal{M}}_{\text{O}}$) contributions,  as ${\cal{M}} = {\cal{M}}_{\text{SM}} + {\cal{M}}_{\text{O}}$. The behaviour of partonic cross sections is then given by
\begin{equation}
{\text{d}\sigma \over  \text{dLIPS}}\sim  | {\cal{M}}_\text{SM} |^2 + 2\text{Re}({\cal{M}}_\text{SM} {\cal{M}}_{\text{O}}^\ast) +  | {\cal{M}}_{\text{O}} |^2\,.
\label{eq:expand}
\end{equation}
The first and third terms proportional to the squared values of the couplings probe CP-even aspects of \hbox{(multi-)Higgs} production, such as cross sections and transverse momentum distributions. The contribution of CP-odd couplings to the interference term has a net-zero effect for CP-even observables. These cancellations can be resolved through the use of specifically tailored CP-odd observables. To explore the CP sensitivity of these operators, it is often advantageous to create ``signed" observables that are responsive to the CP-violating term in the amplitude~\cite{Atwood:1991ka,Plehn:2001nj,Goncalves:2018agy,Gritsan:2020pib,Barman:2021yfh,Bhardwaj:2021ujv}. However, it is important to note that in scenarios where statistical data is limited, obtaining a binned distribution may not always be feasible. This limitation can persist even during the high-luminosity phase of the Large Hadron Collider for certain processes, which is our assumption for di-Higgs WBF production in the following.

\begin{figure*}[t!]
   \centering
    \subfigure[]{\includegraphics[width=0.45\textwidth]{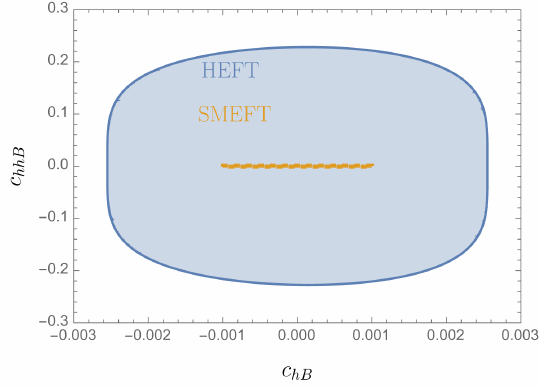}}
     \subfigure[]{\includegraphics[width=0.45\textwidth]{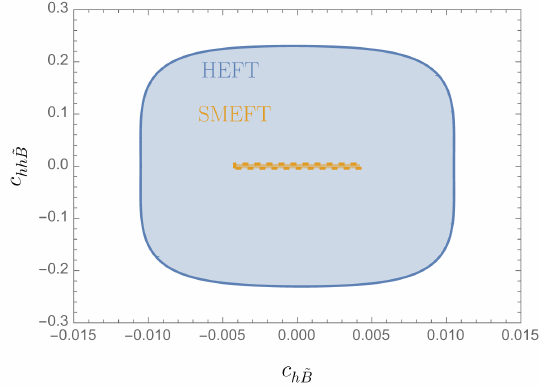}}
    \subfigure[]{\includegraphics[width=0.45\textwidth]{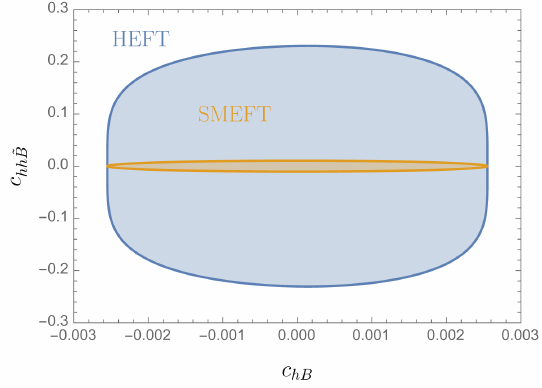}}
     \subfigure[]{\includegraphics[width=0.45\textwidth]{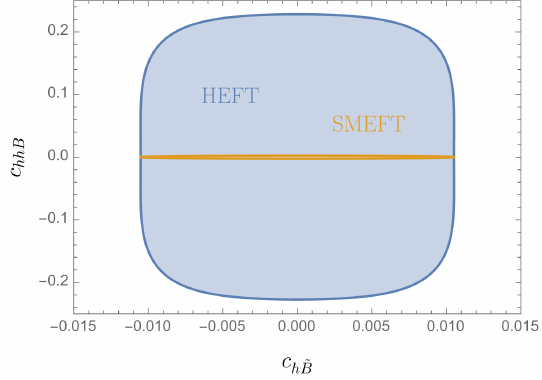}}
    \caption{$95\%$ confidence level regions for the Wilson coefficients: (a)  $(c_{hB},c_{hhB})$, (b) $(c_{h\tilde{B}},c_{hh\tilde{B}})$, (c)  $(c_{hB},c_{hh\tilde{B}})$, and (d)~$(c_{h\tilde{B}},c_{hhB})$ in the HEFT (blue) and SMEFT (orange), similar to Fig.~\ref{fig:HEFTcorW}. To aid in the comparison between HEFT and SMEFT, we set $\Lambda=v$. We consider the HL-LHC with 3~ab$^{-1}$ of data.}
        \label{fig:HEFTcorB}
\end{figure*}

\section{Processes and Analysis}
\label{sec:analysis}
\subsubsection*{Single Higgs production}
\noindent The recent results from the ATLAS and CMS experiments constrain the CP properties of the Higgs boson and its anomalous couplings with electroweak gauge bosons. Nonetheless, it is still possible to accommodate small BSM modifications to them. The tensor structure of the $VVh$ coupling is primarily probed through single Higgs production via WBF, associated production with a weak vector boson ($Zh/Wh$), and Higgs decay to a pair of gauge bosons. The current expected constraints on SMEFT Wilson coefficients in the Higgs-gauge sector from CMS, using $137~\text{fb}^{-1}$ of data, are~\cite{CMS:2021nnc}
\begin{align}
c_{\Phi B}&=[-0.08,0.03]\,,             &c_{\Phi \widetilde{B}}&=[-0.33, 0.33]\,, \nonumber\\
c_{\Phi W}&=[-0.28,0.39]\,,           &c_{\Phi \widetilde{W}}&=[-1.11, 1.11]   \,,
\label{eq:ciCMS}\\
c_{\Phi WB}&=[-0.31, 0.42]\,,&c_{\Phi \widetilde{W}B}&=[-1.21, 1.21]\,.\nonumber
\end{align}

\subsubsection*{Probing non-linearity: $hhjj$ production}
\noindent The WBF channel for Higgs boson pair production is recognized as the second most prevalent process within the SM framework for $hh$ production. It holds a distinctive significance as it is the main channel for probing the quartic gauge-Higgs contact interaction $VVhh$. This uniquely positions WBF di-Higgs production as the primary candidate for probing potential non-linearities of the gauge-Higgs sector, providing critical information on the mechanism underlying electroweak symmetry breaking~\cite{Alonso:2021rac,Englert:2023uug}.
To estimate the sensitivity in particular to CP-odd interactions beyond that, we consider the analysis by the ATLAS collaboration of Ref.~\cite{ATLAS:2022jtk}, which constrains the signal strength as 
\begin{equation}
\label{eq:vbfhh}
{\sigma_\text{WBF}({hh}) \over \sigma_\text{WBF}^{\text{SM}}({hh})}< 7.5 \,,
\end{equation}
at $95\%$ confidence level (CL). This corresponds to the bound $ 0.1 < 1+\zeta_2 < 2.0 $, where $\zeta_2$ is the parameter for the quartic interaction vertex $VVhh$.
We can use this result for $1+\zeta_2$ alongside the SM cross section to estimate the statistical and systematic uncertainties relevant for the constraint of Ref.~\cite{ATLAS:2022jtk} at an integrated luminosity of $137~\text{fb}^{-1}$. We then extrapolate these uncertainties for the HL-LHC with a target luminosity of $3~\text{ab}^{-1}$ using a conservative~\cite{Belvedere:2024wzg} ${\sqrt{\text{luminosity}}}$ scaling. We follow a similar approach for the HL-LHC projection of the Wilson coefficients of Eq.~\eqref{eq:ciCMS}. To approximate the statistical and systematic uncertainties, we construct the $\chi^2$ as the sum over all bins of the squared differences between the observed event counts $N_i$ and the corresponding counts $N_{i}^\text{SM}$ predicted by the SM, divided by the squared uncertainties $\sigma^2_i$ associated with each bin. The $\chi^2$ statistic is expressed as
\begin{equation}
\chi^2 = \sum_{i} \frac{(N_i - N^{\text{SM}}_i)^2}{\sigma_{i,\text{syst}}^2 + \sigma_{i,\text{stat}}^2}\,.
 \label{eq:chi_vvhh}
\end{equation}
These estimates then give rise to HEFT and SMEFT-allowed parameter space from single and double Higgs constraints. Here, $i$ runs over the rate information of the different single and double Higgs channels.

\subsubsection*{A toy fit of non-linear gauge-Higgs CP violation}
To constrain the parameters introduced in Sec.~\ref{sec:corr}, we implement the HEFT Lagrangian in {\tt{FeynRules}}~\cite{Christensen:2008py}, generating a UFO model file~\cite{Degrande:2011ua}, and subsequently interfaced with {\tt{MadGraph5\_aMC@NLO}}~\cite{Frixione:2002ik,Hoeche:2011fd}. We use {\tt{MadGraph5\_aMC@NLO}} to perform the interpolation for the cross section of the $hh$ process in WBF production, studying the correlation between pairs of EFT coefficients. In this analysis, the total rate can be parametrized as a power series involving six template contributions of reference new physics coupling choices in addition to the SM expectation (see also~\cite{ATLAS:2023qzf}). If an EFT coefficient is CP-odd, its linear contribution vanishes from the CP-even cross section interpolation, as mentioned above. 

\begin{figure}[!t]
   \centering
   \includegraphics[width=0.5\textwidth]{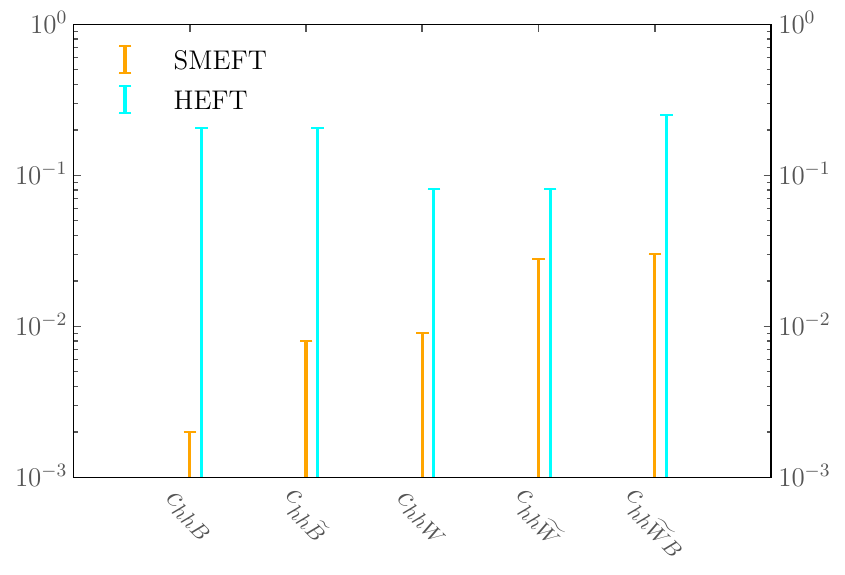}
       \caption{$95\%$ CL limits for the HEFT and SMEFT Wilson coefficients from the HL-LHC extrapolation to 3/ab. We assume $\Lambda=v$ for comparability. As single Higgs results include jet-based asymmetries, the SMEFT constraints between CP-even and CP-odd results are not symmetric.}
    \label{fig:dihiggs}
\end{figure}

Reference~\cite{ATLAS:2022jtk} considers $hh$ production via WBF with subsequent Higgs boson decay to bottom quark pairs $h\to b\bar{b}$. Higher-dimensional gauge-Higgs interactions can modify the partial decay width of the Higgs decays into vector bosons $h \to VV^{*}$, which indirectly affect the dominant $h \to$ $b\bar{b}$ branching ratio. The contributions of the considered HEFT and SMEFT operators to the Higgs decay widths, compared to those in the SM, are detailed in Appendix~\ref{app:dw-appendix}. Our analysis includes these correlated effects that propagate to the exclusive $4b$ final state.

We can then obtain $95\%$~CL regions on the HEFT and SMEFT parameters as pair-wise combinations of the EFT coefficients for $3~\text{ab}^{-1}$ luminosity. By combining the bounds from the WBF $hh$ and single Higgs analyses~\cite{CMS:2021nnc}, we calculate a total $\chi^2$ and identify the 95\% exclusion depending on the relevant number of degrees of freedom. In Fig.~\ref{fig:HEFTcorW}, we show the correlations across different Higgs multiplicities for HEFT and SMEFT parameters. In the HEFT framework, the single and double Higgs operators are completely independent. Therefore, the bounds obtained in this framework for the EFT coefficients $c_{hhW}$ and $c_{hh\widetilde{W}}$ are solely determined by the $hh$ WBF analysis, even when the single Higgs constraints are reflected in the limit setting, as detailed above. 

In contrast, the correlations between different Higgs multiplicities shown in Fig.~\ref{fig:HEFTcorW} are fixed in the SMEFT framework. This is because both single and double Higgs interactions are governed by the same high-dimensional operators in SMEFT, \emph{i.e.},  $c_{hhW}= c_{hW}$ and $c_{hh\widetilde{W}}= c_{h\widetilde{W}}$. Hence, for the SMEFT, the WBF di-Higgs measurements do not provide additional phenomenological relevance beyond single Higgs observations for the interactions considered. Furthermore, in both HEFT and SMEFT, $c_{hW}$ and $c_{h\widetilde{W}}$ are primarily constrained from the single Higgs measurements and thus are stringently constrained. We also observe that the SMEFT contours in the top panels (a) and (b) of Fig.~\ref{fig:HEFTcorW} are more constrained than those in the bottom panels (c) and (d). This is because the two-dimensional correlations in the SMEFT shown in the top panel arise from the same SMEFT operator, reducing the number of degrees of freedom in the $\chi^2$ fit.  For completeness, we present in Fig.~\ref{fig:HEFTcorB} the corresponding limits on the correlations involving the $B$ field.

In Fig.~\ref{fig:dihiggs}, we show the summary plot of the 95\% CL limits for the EFT coefficients involving the double Higgs interactions. Although the production rate of double Higgs in the WBF channel is small compared to single Higgs production, resulting in relatively weak limits, we find that it still provides useful constraints on potential non-linearities of the gauge-Higgs sector. 

We note that the main limiting factor of our extrapolation is the as-yet unclear performance improvement for WBF $hh$ production. These are known to be large, albeit process-dependent~\cite{Belvedere:2024wzg}. A luminosity-extrapolated constraint should therefore be considered conservative. An immediate consequence of this is that a marginalised study including all relevant interactions would lead to no notable sensitivity. Our results show, however, that 10\% deviations in the gauge-Higgs sector at the weak scale should be attainable, which could elucidate potential non-linear modifications of the gauge-Higgs sector (or a lack thereof).

\section{Conclusions}
\label{sec:conclusion}
The search for new physics beyond the Standard Model is well underway at the LHC and with more data becoming available towards the high luminosity phase, rare processes such as WBF Higgs pair production can be probed with increasing scrutiny. On the one hand, when the physical Higgs boson is considered to arise as part of a weak doublet, the relevance of these channels is suppressed by gauge symmetry arguments. However, these correlations can and are being assessed using {\emph{actual}} LHC data~\cite{ATLAS:2023qzf,CMS:2022hgz}. From a theoretical perspective, large deviations are possible if electroweak symmetry breaking significantly departs from the SM expectation~\cite{Alonso:2021rac}. This still falls within the phenomenologically allowed coupling patterns observed for the Higgs boson~\cite{Englert:2023uug} in the current stage of the LHC Higgs physics program. As traditional BSM scenarios are equally challenged by the LHC's results so far, it seems prudent to also consider coupling modifications away from SMEFT choices in parallel.

To this end, in this work, we have considered the relation of SMEFT and HEFT for the tell-tale WBF di-Higgs process, in particular from a perspective of CP-violation in the gauge-Higgs sector. It is not clear yet whether jet-differential information will be available for this final state, which would isolate genuine CP-odd effects. However, the CP-even rate information available already~\cite{ATLAS:2023qzf,CMS:2022hgz} can be used to indirectly constrain CP-odd multi-Higgs gauge boson interactions and their correlation across different Higgs multiplicities. When considering physics that follows the SMEFT pattern, WBF di-Higgs analyses do not necessarily provide stringent additional constraints. Contrary to that, HEFT opens up an entirely new territory, extending to CP-odd modifications of the Higgs gauge sector in this mode. Our results therefore provide additional motivation to further consider the strategies of~\cite{ATLAS:2023qzf,CMS:2022hgz}, including their extension to CP-odd interactions.

\acknowledgments
AB, DG and AN thank the U.S.~Department of Energy for the financial support, under grant number DE-SC 0016013. CE is supported by the UK Science and Technology Facilities Council (STFC) under grant ST/X000605/1, the Leverhulme Trust under RPG-2021-031 and the Insitute for Particle Physics Phenomenology Associate Scheme.

\appendix
\section{Higgs boson decay widths}
\label{app:dw-appendix}
In this appendix, we present the Higgs decay widths for $h\to \gamma\gamma,Z\gamma, ZZ^*~\text{and}~WW^*$ in the presence of the gauge-Higgs operator for HEFT and SMEFT parametrizations relative to the SM results. As discussed in Sec.~\ref{sec:corr}, we set $c_{\Phi WB}=0$ and $c_{h WB}=0$ throughout due to the stringent constraints imposed by the $S$ parameter:
\begin{widetext}
\begin{align}
\frac{\Gamma_{\text{HEFT}}(h \to \gamma \gamma)}{\Gamma_{\text{SM}}(h \to \gamma \gamma)} = &
1 + \frac{840.243 \, c_{hB}}{(v/\text{246 GeV})} 
+ \frac{239.296 \, c_{hW}}{(v/\text{ 246 GeV})} + \frac{176502 \, c_{hB}^2}{{(v/\text{246 GeV})}^2} + \frac{176502 \, c_{h\widetilde{B}}^2}{{(v/\text{246 GeV})}^2} - \frac{188513 \, c_{h\widetilde{B}} \, c_{h\widetilde{W}B}}{{(v/\text{246 GeV})}^2} \nonumber \\
&+ \frac{50266.7 \, c_{h\widetilde{W}B}^2}{{(v/\text{246 GeV})}^2} + \frac{100533 \, c_{hB} \, c_{hW}}{{(v/\text{246 GeV})}^2} 
+ \frac{14315.7 \, c_{hW}^2}{{(v/\text{246 GeV})}^2} + \frac{100533 \, c_{h\widetilde{B}} \, c_{h\widetilde{W}}}{{(v/\text{246 GeV})}^2} 
\nonumber \\
&- \frac{53687.3 \, c_{h\widetilde{W}B} \, c_{h\widetilde{W}}}{{(v/\text{246 GeV})}^2} + \frac{14315.7 \, c_{h\widetilde{W}}^2}{{(v/\text{246 GeV})}^2}\,, \\
%
%
\frac{\Gamma_{\text{HEFT}}(h \to \gamma Z)}{\Gamma_{\text{SM}}(h \to \gamma Z)} = &
1 + \frac{4798.25 \, c_{hB}}{{(v/\text{246 GeV})}} 
- \frac{4798.25 \, c_{hW}}{{(v/\text{ 246 GeV})}} 
+ \frac{578701 \, c_{hB}^2}{{(v/\text{246 GeV})}^2} + \frac{578701 \, c_{h\widetilde{B}}^2}{{(v/\text{246 GeV})}^2} 
- \frac{777991 \, c_{h\widetilde{B}} \, c_{h\widetilde{W}B}}{{(v/\text{246 GeV})}^2} 
\nonumber \\
&+ \frac{261478 \, c_{h\widetilde{W}B}^2}{{(v/\text{246 GeV})}^2} - \frac{1.1574 \times 10^{6} \, c_{hB} \, c_{hW}}{{(v/\text{246 GeV})}^2} 
+ \frac{578701 \, c_{hW}^2}{{(v/\text{246 GeV})}^2} - \frac{1.1574 \times 10^{6} \, c_{h\widetilde{B}} \, c_{h\widetilde{W}}}{{(v/\text{246 GeV})}^2} \nonumber \\
&+ \frac{777991 \, c_{h\widetilde{W}B} \, c_{h\widetilde{W}}}{{(v/\text{246 GeV})}^2} + \frac{578701 \, c_{h\widetilde{W}}^2}{{(v/\text{246 GeV})}^2}\,,\\
\frac{\Gamma_{\text{HEFT}}(h \to ZZ^{*})}{\Gamma_{\text{SM}}(h \to ZZ^{*})} = &
1 - \frac{9.2 \times 10^{-5} \, c_{hB}}{(v/\text{246 GeV})} 
- \frac{2.9 \times 10^{-6} \, c_{hW}}{(v/\text{ 246 GeV})} - \frac{0.200 \, c_{hB}^2}{{(v/\text{246 GeV})}^2} + \frac{0.200 \, c_{h\widetilde{B}}^2}{{(v/\text{246 GeV})}^2} + \frac{0.057 \, c_{h\widetilde{B}} \, c_{h\widetilde{W}B}}{{(v/\text{246 GeV})}^2} 
\nonumber \\
&- \frac{0.243\, c_{h\widetilde{W}B}^2}{{(v/\text{246 GeV})}^2} - \frac{0.0872 \, c_{hB} \, c_{hW}}{{(v/\text{246 GeV})}^2} 
- \frac{0.198\, c_{hW}^2}{{(v/\text{246 GeV})}^2} - \frac{0.0872 \, c_{h\widetilde{B}} \, c_{h\widetilde{W}}}{{(v/\text{246 GeV})}^2} \nonumber \\
&
+ \frac{2.9 \times 10^{-6} \, c_{h\widetilde{W}B} \, c_{h\widetilde{W}}}{{(v/\text{246 GeV})}^2} 
- \frac{0.198 \, c_{h\widetilde{W}}^2}{{(v/\text{246 GeV})}^2}\,, \\
%
%
\frac{\Gamma_{\text{HEFT}}(h \to WW^{*})}{\Gamma_{\text{SM}}(h \to WW^{*})} = &
1 + \frac{1.3 \times 10^{-7} \, c_{hB}}{{(v/\text{246 GeV})}} 
- \frac{0.0013 \, c_{hW}}{{(v/\text{ 246 GeV})}} - \frac{0.243 \, c_{hB}^2}{{(v/\text{246 GeV})}^2} - \frac{0.243 \, c_{h\widetilde{B}}^2}{{(v/\text{246 GeV})}^2} 
- \frac{1.3 \times 10^{-7}\, c_{h\widetilde{B}} \, c_{h\widetilde{W}B}}{{(v/\text{246 GeV})}^2} 
\nonumber \\
&- \frac{0.243 \, c_{h\widetilde{W}B}^2}{{(v/\text{246 GeV})}^2} - \frac{6.24 \times 10^{-8} \, c_{hB} \, c_{hW}}{{(v/\text{246 GeV})}^2} 
- \frac{0.243 \, c_{hW}^2}{{(v/\text{246 GeV})}^2} - \frac{6.24 \times 10^{-8} \, c_{h\widetilde{B}} \, c_{h\widetilde{W}}}{{(v/\text{246 GeV})}^2} \nonumber \\
&+ \frac{4.38 \times 10^{-7} \, c_{h\widetilde{W}B} \, c_{h\widetilde{W}}}{{(v/\text{246 GeV})}^2} - \frac{0.0243\, c_{h\widetilde{W}}^2}{{(v/\text{246 GeV})}^2}\,,
\end{align}
\end{widetext}
\begin{widetext}
\noindent where $v$ is understood as the interaction scale of the isosinglet Higgs boson. For the SMEFT case these are 
\begin{align}
\frac{\Gamma_{\text{SMEFT}}(h \to \gamma \gamma)}{\Gamma_{\text{SM}}(h \to \gamma \gamma)} =& 
1 + \frac{50.84\, c_{\Phi B}}{(\Lambda/\text{1 TeV})^2} 
- \frac{14.48 \, c_{\Phi W}}{(\Lambda/\text{1 TeV})^2} + \frac{646.38 \, c_{\Phi B}^2}{(\Lambda/\text{1 TeV})^4} + \frac{646.38 \, c_{\Phi \widetilde{B}}^2}{(\Lambda/\text{1 TeV})^4} 
- \frac{690.36 \, c_{\Phi \widetilde{B}} \, c_{\Phi \widetilde{W}B}}{(\Lambda/\text{1 TeV})^4}
 \nonumber \\
&+ \frac{184.08 \, c_{\Phi \widetilde{W}B}^2}{(\Lambda/\text{1 TeV})^4} - \frac{ 368.17 \, c_{\Phi B} \, c_{\Phi W}}{(\Lambda/\text{1 TeV})^4}
+ \frac{52.42 \, c_{\Phi W}^2}{(\Lambda/\text{1 TeV})^4} - \frac{ 368.17 \, c_{\Phi \widetilde{B}} \, c_{\Phi \widetilde{W}}}{(\Lambda/\text{1 TeV})^4} \nonumber \\
&+ \frac{196.61 \, c_{\Phi \widetilde{W}B} \, c_{\Phi \widetilde{W}}}{(\Lambda/\text{1 TeV})^4} + \frac{ 52.42 \, c_{\Phi \widetilde{W}}^2}{(\Lambda/\text{1 TeV})^4}\,,
\\
\frac{\Gamma_{\text{SMEFT}}(h \to \gamma Z)}{\Gamma_{\text{SM}}(h \to \gamma Z)} =& 
1 + \frac{290.371 \, c_{\Phi B}}{(\Lambda/\text{1 TeV})^2} 
- \frac{290.371 \, c_{\Phi W}}{(\Lambda/\text{1 TeV})^2} + \frac{2119.31 \, c_{\Phi B}^2}{(\Lambda/\text{1 TeV})^4} + \frac{2119.31 \, c_{\Phi \widetilde{B}}^2}{(\Lambda/\text{1 TeV})^4} 
- \frac{2849.15 \, c_{\Phi \widetilde{B}} \, c_{\Phi \widetilde{W}B}}{(\Lambda/\text{1 TeV})^4}
 \nonumber \\
&+ \frac{957.581 \, c_{\Phi \widetilde{W}B}^2}{(\Lambda/\text{1 TeV})^4} - \frac{ 4238.62 \, c_{\Phi B} \, c_{\Phi W}}{(\Lambda/\text{1 TeV})^4}
+ \frac{2119.31 \, c_{\Phi W}^2}{(\Lambda/\text{1 TeV})^4} - \frac{ 4238.62 \, c_{\Phi \widetilde{B}} \, c_{\Phi \widetilde{W}}}{(\Lambda/\text{1 TeV})^4} \nonumber \\
&+ \frac{2849.15 \, c_{\Phi\widetilde{W}B} \, c_{\Phi \widetilde{W}}}{(\Lambda/\text{1 TeV})^4} + \frac{ 2119.31 \, c_{\Phi \widetilde{W}}^2}{(\Lambda/\text{1 TeV})^4}\,,
\end{align}
\end{widetext}
whilst the $WW,ZZ$ partial decays can be obtained directly from the HEFT parameterization using identifications like~Eqs.~\eqref{eq:smeftheft} and~\eqref{eq:smeftheft2} (which extend to the other operators discussed in this work).
\bibliographystyle{JHEP}
\bibliography{references} 
\end{document}